\documentstyle[12pt]{article}
\input epsf
\begin{document}
\title{Comparison of the fits to data on polarised structure functions
and spin asymmetries}
\author{ Jan
Bartelski\\ Institute of Theoretical Physics, Warsaw University,\\
Ho$\dot{z}$a 69, 00-681 Warsaw, Poland. \\ \\ \and Stanis\l aw
Tatur
\\ Nicolaus Copernicus Astronomical Center,\\ Polish Academy of
Sciences,\\ Bartycka 18, 00-716 Warsaw, Poland. \\ }
\date{}
\maketitle
\vspace{1cm}
\begin{abstract}
\noindent In order to obtain polarised parton densities we have
made next to leading order QCD fit using experimental data on deep
inelastic structure functions on nucleons. This fit is compared
with the updated fit to corresponding spin asymmetries. We get
very similar results for all fits also for different data samples.
It seems that only polarised parton densities for non-strange
quarks $\Delta u$ and $\Delta d$ are relatively well determined
from the present polarised deep inelastic experiments. Integrated
gluon contribution at  $ Q^{2}=1\, {\rm GeV^{2}}$ is, as in our
previous fits, very small.
\end{abstract}

\newpage
In order to get polarised parton (i.e. quark and gluon) densities
one uses data on deep inelastic scattering of polarised lepton on
polarised nucleon targets.  Quite a lot of data exist for such
scattering  on different nucleon targets. The data come from
experiments made at SLAC [1-10],  CERN [11-16] and DESY
\cite{hermes,hernew}. Recently the data from SLAC from E155
\cite{e155p} experiment on protons has been published. The
experimental groups present data for spin asymmetries as well as
for polarised structure functions.

The analysis of the EMC group results \cite{emc} started an
interest in studying such data. So called "spin crisis" was
connected with the fact that only very little of spin of nucleon
was carried by quarks. The suggestion from ref. \cite{dg} was that
polarised gluons may be responsible for this phenomena. Many next
to leading (NLO) order QCD analysis [20-27] were performed  and
polarised parton distributions  were determined. The main purpose
of this paper is to use data for polarised structure functions
$g_{1}(x,Q^2)$ on proton, neutron and deuteron targets in order to
determine polarised parton distributions. This fit will be
compared with our updated (in which we take into account recently
published data on protons from E155 \cite{e155p} experiment in
SLAC) fits to spin asymmetries. It was advocated by us \cite{bt1}
and \cite{grv} that using spin asymmetries for determination of
polarised parton densities one avoids the problem with higher
twist contributions.
 On the other hand
it is the polarised structure fuctions and polarised parton
distributions that we want to determine.

 We  compare these two ways of making fits with similar technical
 assumptions and the
same parton functions used for fitting. We will see that both
methods give very similar results for parton distributions. Our
previous \cite{BT,BTn} important conclusion that the integrated
gluon contribution is negligible at $ Q^{2}=1\, {\rm GeV^{2}}$
does not change. Most of the groups used experimental data for
spin asymmetries  to determine polarised parton distributions. In
addition to spin asymmetries one has also experimental data for
polarised structure functions calculated from the spin asymmetries
in a specific way chosen by an experimental group. There were also
several fits to the data on polarised structure functions
\cite{alt0,SMCth}. As in \cite{BT,BTn} we will make fits to  two
samples of the data. In the first group we will have  data  for
the same $x$ (strictly speaking for the near values) and different
$Q^2$ and in the second the "averaged" data where one averages
over $Q^2$ (the errors are smaller and $Q^2$ dependence is smeared
out). In most of the fits to experimental data only second sample
(namely with averaged $Q^2$ dependence) was used. Our fits use the
both sets of data. The data for polarised structure functions are
usually given only for averaged sample of data (the exception is
E155 experiment for deuteron and proton data).

Experiments on unpolarised targets provide information on the
unpolarised quark densities $q(x,Q^2)$ and $G(x,Q^2)$ inside the
nucleon. These densities can be expressed in term of
$q^{\pm}(x,Q^2)$ and $G^{\pm}(x,Q^2)$, i.e. densities of quarks
and gluons with helicity along or opposite to the helicity of the
parent nucleon.
\begin{equation}
q = q^{+}+q^{-},\hspace*{1.5cm} G = G^{+}+G^{-}.
\end{equation}
$q$ stands for quark and antiquark contributions.

The polarised parton densities, i.e. the differences of $q^{+}$,
$q^{-}$ and $G^{+}$, $G^{-}$ are given by:
\begin{equation}
\Delta q = q^{+}-q^{-},\hspace*{1.5cm} \Delta G = G^{+}-G^{-}.
\end{equation}

We will try to determine $q^{\pm} (x,Q^2)$ and $G^{\pm} (x,Q^2)$,
in other words, we will try to connect unpolarised and polarised
data.

In our fits we will use functions for the polarised parton
densities that are suggested by the fit to unpolarised data. We
risk that asymptotic behaviour of our parton distributions is not
correct but it seems to us not so important when we limit
ourselves to the measured region of x. Maybe it is also not bad to
use parametrisation completely different from other groups to
check how the results depend on that. Our starting point are the
formulas for unpolarised quark and gluon distributions gotten (at
$ Q^{2}=1\, {\rm GeV^{2}}$) from the fit performed by Martin,
Roberts, Stirling and Thorne \cite{MRSTnew} (they use
$\Lambda^{n_{f}=4}_{\overline{MS}}=0.3$ $\mbox{GeV}$ and
$\alpha_s(M^2_Z)=0.120$)

We will not use small and high x behaviour of unpolarised parton
distributions as fitted parameters as some other groups do. We
will split $q$ and $G$, as was already discussed in ref.
\cite{BT,BTn}, into two parts in such a manner that the distributions
$q^{\pm}(x,Q^2)$ and $G^{\pm}(x,Q^2)$ remain positive. Our
polarised densities  for quarks  and gluons are parametrised as
follows:

\begin{eqnarray}
\noindent \Delta
u_{v}(x)=x^{-0.5911}(1-x)^{3.395}(a_{1}+a_{2}\sqrt {x}+a_{4}x),
\nonumber \\ \noindent \Delta
d_{v}(x)=x^{-0.7118}(1-x)^{3.874}(b_{1}+b_{2}\sqrt{x}+b_{3}x),
\nonumber \\ \noindent 2\Delta \bar{u} (x)=0.4\Delta M(x)-\Delta
\delta (x), \\ \noindent 2\Delta \bar{d} (x)=0.4\Delta M(x)+\Delta
\delta (x), \nonumber
\\
\noindent  2\Delta \bar{s} (x)=0.2\Delta M_s(x), \nonumber \\
\noindent \Delta G (x)=x^{-0.0829}(1-x)^{6.587}(d_1+d_2
\sqrt{x}+d_3 x). \nonumber
\end{eqnarray}

\noindent whereas for the antiquarks ( and sea quarks ):
\begin{eqnarray}
\noindent \Delta
M(x)=x^{-0.7712}(1-x)^{7.808}(c_{1}+c_{2}\sqrt{x}), \nonumber \\
\noindent \Delta M_s= x^{-0.7712}(1-x)^{7.808}
(c_{1s}+c_{2s}\sqrt{x}),  \\ \noindent \Delta \delta
(x)=x^{0.183}(1-x)^{9.808}c_{3}(1+9.987x-33.34x^{2}), \nonumber
\end{eqnarray}

We also have

\begin{eqnarray}
\noindent \Delta u=\Delta u_{v}+2 \Delta \bar{u} , \nonumber \\
\noindent \Delta d=\Delta d_{v}+2 \Delta \bar{d} ,  \\ \noindent
\Delta s=2 \Delta \bar{s}, \nonumber
\end{eqnarray}
 We use as before \cite{BT,BTn} additional independent parameters
for the strange sea contribution with the same as for non-strange
sea functional dependence. Maybe not all parameters are important
in the fit and it could happen that some of the coefficients in
eq.(3,4) taken as free parameters in the fit are small or in some
sense superfluous. Putting  them to zero (or eliminating them)
increase $\chi^{2}$ only a little but makes  the quantity
$\chi^{2}/N_{DF}$ smaller. We will see that that is the case with
some parameters introduced in eq. (3,4).

In order to get the unknown parameters in the expressions for
polarised quark and gluon distributions (eqs.(3,4)) we calculate
the spin asymmetries (starting from initial $Q^2$ = 1
$\mbox{GeV}^2$) for measured values of $Q^2$ and make a fit to the
experimental data on spin asymmetries for proton, neutron and
deuteron targets. The spin asymmetry $A_1(x,Q^2)$ can be expressed
via the polarised structure function $g_1(x,Q^2)$ as
\begin{equation}
A_1(x,Q^2)\cong \frac{(1+\gamma^{2}) g_{1}(x,Q^2)}{ F_1(x,Q^2)}=
\frac{ g_{1}(x,Q^2)}{ F_2(x,Q^2)}[2x(1+ R(x,Q^2))],
\end{equation}
\noindent where  $R = [F_2(1+\gamma^{2})-2xF_1]/ 2xF_1$ whereas
$F_1$ and $F_2$ are the unpolarised structure functions and
$\gamma =2Mx/Q$ ($M$ stands for proton mass). We will take the new
determined value of $R$ from the \cite{whitn}. The factor
$(1+\gamma^{2})$ plays non negligible role for $x$ and $Q^{2}$
values measured in SLAC experiments. In calculating $g_{1}(x,Q^2)$
and $F_{2}(x,Q^2)$ in the next to leading order we use procedure
described in \cite{BT} following the method described in
\cite{grv,ewol2} performing calculations with Mellin transforms
and then calculating Mellin inverse.
 Having calculated the
asymmetries according to equation (6) for the  value of $Q^2$
obtained in experiments we can make a fit to asymmetries on
proton, neutron and deuteron targets. The other possibility is to
use directly the data for the polarised structure function
$g_{1}(x,Q^2)$ (the values of $g_1$ were given for the averaged
values of $Q^2$) on proton, neutron and deuteron targets to
determine unknown coefficients in expressions for polarised parton
distributions. The problem is that different experimental groups
used their own specific methods to obtain the values of polarised
structure functions $g_{1}(x,Q^2)$ from the measured asymmetries.
At the end we want to know polarised structure functions and
polarised parton distributions. In order to calculate them from
spin asymmetries we have to choose what shall we take for
$F_{1}(x,Q^2)$ or $F_{2}(x,Q^2)$ and $R(x,Q^2)$. As was already
mentioned we calculated $F_{2}(x,Q^2)$ in NLO for actual values of
$x$ and $Q^2$ using quark and gluon contributions for $ Q^{2}=1\,
{\rm GeV^{2}}$ given by MRST \cite{MRSTnew}. The values of R were in the
earlier fits taken from Whitlow \cite{whit} and later from E143 group \cite{whitn}.We
have treated all  experiments in the same way. There is also a
problem of higher twist corrections (power low corrections to R
were included). We will not include the higher twist corrections
because of still big experimental errors. The spread of the
results could be a measure of uncertainties in both methods. We
will compare fits using determination of parameters from polarised
structure functions and from spin asymmetries. As will be seen
later the results obtained by two methods are very similar.

It was already seen from our previous paper \cite{BTn} that $g_{1}(x,Q^2)$
calculated from spin asymmetries fits not bad the data points for
$g_1$. Data points for polarised structure functions were given
for averaged data set so it is natural to compare fit to $g_1$
(called by us fit $g$) with the fit to spin asymmetries (fit
$A_{1}$) with the same number of points. We will take into account
in this case 197 data points (We will take E155 proton and
deuteron data without averaging). These fits will be compared with
the fit (fit $A_{2}$) to spin asymmetries for non averaged data
where we take into account 465 experimental data points. At the
beginning we will not put any constrains from hyperon decays.
Later we will also not fix $a_8=\Delta u+\Delta d-2 \Delta s$
value but we will add experimental point $a_8 = 0.58 \pm 0.1$ with
enhanced (to 3$\sigma$) error. That means we will simply add to
$\chi^{2}$ corresponding to experimental points for spin
asymmetries the term connected with experimental point from
hyperon decays. We will discuss how this additional experimental
point influences our results.

Not all parameters are important in the fits.  It seems that some
of the parameters of the most singular terms are superfluous and
we can eliminate them. We will put $d_{1}=0$ (such assumption
gives that $\delta G/G \sim x^{1/2}$ for small $x$), $b_{1}=0$
(the most singular term in $\Delta d_{v}$) and assume
$c_{1s}=c_{1}$ (i.e. the most singular terms for strange and
non-strange sea contributions are equal). Fixing these four
parameters in the fit practically does not change the value of
$\chi^{2}$ but improves $\chi^{2}/N_{DF}$. We also have to make
some remarks about parameters $c_3$ and $d_2$. In the fit for
$g_1$ these parameters are important i.e. when we eliminate them
$\chi ^2$ per degree of freedom increases. That is not the case in
fits for spin asymmetries. In this case these parameters are
superficial (that means for example that splitting
$\bar{u}-\bar{d}$ is not well defined by data to spin asymmetries.
We will leave this parameters for comparison with fit to $g_1$ (do
not eliminate them) but they could be not well determined and
cause some artificial shifts in other parameters.

\vspace{0.5cm}
Table 1. {\em{The parameters of our three fits
calculated at $ Q^{2}=1\, {\rm  GeV^{2}}$ together with $\chi^2$
per degree of freedom}}

\vspace{0.5cm}

\hspace{-1.5cm}
\begin{tabular}{||c|c|c|c|c|c|c||} \hline
fit &$a_1$&$a_2$&$a_4$&$b_2$&$b_3$&$c_1$
\\ \hline fit $g$&$0.61 $&$-7.05$&$17.1 $&$-2.02$&$0.34$&$-0.34$ \\
 &$\pm 0.0$&$\pm 0.23$&$\pm 0.23$&$\pm 0.0$&$\pm 0.24$&$\pm0.03$ \\ \hline
fit $A_{1}$ &$0.56$&$-5.50$&$14.7$&$-1.67$&$-0.173$&$-0.338$
\\ & $\pm 0.13$&$\pm 1.22$&$\pm 1.66$&$\pm 0.03$&$\pm 0.26$&$\pm
0.10$ \\ \hline fit $A_{2}$&
$0.49$&$-5.34$&$14.66$&$-2.02$&$0.35$&$-0.32$
\\ &$\pm 0.13$&$\pm 1.20$&$\pm 1.63$&$\pm 0.0$&$\pm 0.25$&$\pm
0.10$ \\ \hline
\end{tabular}
\vspace{0.5cm}

\hspace{2cm}
\begin{tabular}{||c|c|c|c|c|c|c||} \hline
fit&$c_{2}$&$c_{2s}$&$c_3$&$d_2$&$d_3$&$\chi^2/N_{DF}$
\\ \hline
fit $g$&$4.15 $&$4.15 $&$-1.05 $&$-29.0$&$87.1$&$0.87$ \\
&$\pm0.0$&$\pm 0.0$&
 $\pm 0.56$&$\pm8.6$&$\pm36.1$& \\ \hline
fit $A_{1}$&$3.23$&$4.15$&$-0.617$&$-15.4$&$42.2$&$0.81$ \\ &$\pm
0.80$&$\pm 0.22$& $\pm 0.58$&$\pm 0.22$&$\pm 15.0$& \\ \hline fit
$A_{2}$&$3.69$&$4.15$&$-0.39$&$-14.0$&$27.0$&$0.84$ \\ &$\pm
0.79$&$\pm 0.20$&
 $\pm 0.48$&$\pm 0.04$&$\pm 11.4$& \\ \hline
 \end{tabular}
 \vspace{0.5cm}

In the Table 1 we present the values of parameter from the fit to
the data on polarised structure functions and spin asymmetries for
averaged and non averaged data together with $\chi^2/N_{DF}$
values.

For example in the case of first row in table 1 corresponding to
the fit to polarised structure functions the obtained quark and
gluon distributions lead for ($Q^2$ =1 $GeV^2$) to the following
integrated (over $x$) quantities: $\Delta u = 0.72 , \Delta d=
-0.64 ,
 \Delta s= 0.05 , \Delta u_v = 0.54 ,
  \Delta d_v = -0.65 , 2\Delta \bar{u} = 0.18
 , 2\Delta \bar{d} = 0.01 .$

We have positively polarised sea for up and down quarks and
positively polarised sea for strange quarks. The gluon
polarisation is small. The value of $a_3=1.36$ was not assumed as
an input in the fit (as is the case in nearly all fits
\cite{inni}) and comes out slightly higher than the experimental
value. The value of $a_8=-0.01$ is completely different from the
experimental figure. Taking into account that fits to polarised
structure functions and spin asymmetries use different methods to
calculate $F_{2}(x,Q^2)$ and $R(x,Q^2)$ there is no reason to
expect that they give exactly the same results. The obtained
values of parameters are very close and practically within
experimental errors. The parameters calculated in fits to spin
asymmetries (line 2 and 3) are closer in comparison with fit to
$g_1$ (line 1) but are not identical. The spread of parameters
measures small differences in the $Q^2$ evolution, differences in
experimental errors and influence of our specific functions used
in fits.

As was already mentioned in \cite{BT} the asymptotic behaviour at
small $x$ of our polarised quark distributions is determined by
the unpolarised ones and hence do not have the expected
theoretically Regge type behaviour. Some of the quantities
specially integrated sea contributions and also some valence
contributions in our fit change rapidly for $x \leq 0.003$. that
is not something that we expect from Regge behaviour with small
exponent.

Hence, we will present quantities integrated over the region from
$x$=0.003 to $x$=1 (it is practically integration over the region
which is covered by the experimental data, except of non
controversial extrapolation for highest $x$).  The values of
integrated quantities in the measured region we consider as more
reliable then those in the whole region.

 The corresponding
quantities for three our  fits  together with $\chi^2/N_{DF}$ are
presented in table 1a.

\vspace{0.5cm}
\newpage
Table 1a. {\em {The values of quark and gluon polarisations  at $
Q^{2}=1\, {\rm
 GeV^{2}}$   for our three fits}}

\vspace{0.5cm}

\begin{tabular}{||c|c|c|c|c|c|c|c|c||} \hline
fit&$\Delta u $&$\Delta d $&$\Delta u_v $&$\Delta d_v $&$2\Delta
\bar{u} $&$2\Delta\bar{d}$&$2\Delta \bar{s} $&$\Delta G$
\\ \hline
fit $g$&$0.74 $&$-0.49$&$0.44 $&$-0.63$&$0.30$&$0.14$&$0.11
$&$0.15 $\\ \hline fit $A_{1}$&
$0.75$&$-0.48$&$0.57$&$-0.57$&$0.18$&$0.09$&$0.11$&$0.01$
\\ \hline
fit $A_{2}$&
$0.76$&$-0.47$&$0.54$&$-0.63$&$0.22$&$0.16$&$0.12$&$-0.19$
\\  \hline
\end{tabular}
\vspace{0.5cm}

 From the table we see that there are changes in
valence and sea contributions in different fits but the values of
$\Delta u$ and $\Delta d$ practically do not differ. We use  the
parametrisation where the most singular term in sea contribution
is very similar to valence quark terms and that maybe is the
reason why splitting into valence and sea contribution is fragile
and changes for different fits but in the case of $\Delta u$ and
$\Delta d$ do not differ much. In the first fit at  $ Q^{2}=1\,
{\rm GeV^{2}}$ we get $\Gamma^{p}_{1}=0.124$,
$\Gamma^{n}_{1}=-0.052$, $a_{3}=1.23$ and $\Delta \Sigma=0.36$
comparing with the second fit to averaged spin asymmetries where
we have $\Gamma^{p}_{1}=0.125$, $\Gamma^{n}_{1}=-0.051$,
$a_{3}=1.24$ and $\Delta \Sigma=0.38$. These results are very
close. The value of $a_{3}$ in the measured region without any
assumption comes out close to the value measured in hyperon
decays.

\begin{figure}
\noindent \hspace{-0.5cm} \epsfxsize =400pt\epsfbox{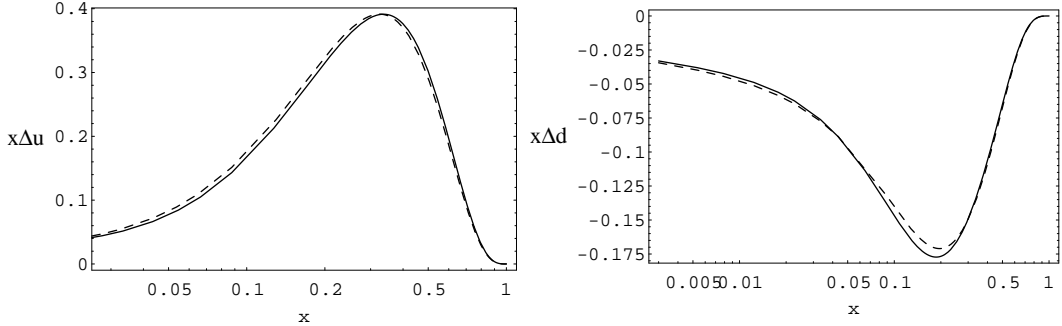}
\caption{\em{The quark and gluon densities for up and down flavour
versus $x$ gotten from the fit $g$ (solid line) and $A_{1}$ (dashed line).}}
\end{figure}

\begin{figure}
\noindent \hspace{-0.5cm} \epsfxsize =400pt\epsfbox{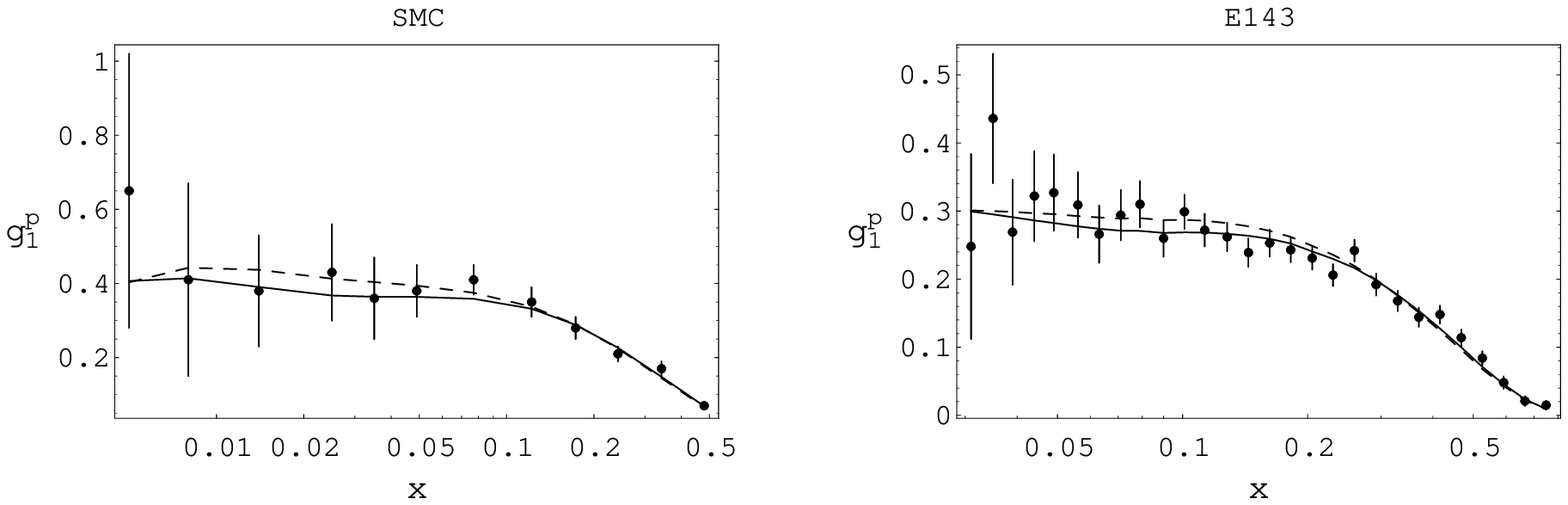}
\vspace{0.5cm}

\noindent \hspace{-0.5cm} \epsfxsize =400pt\epsfbox{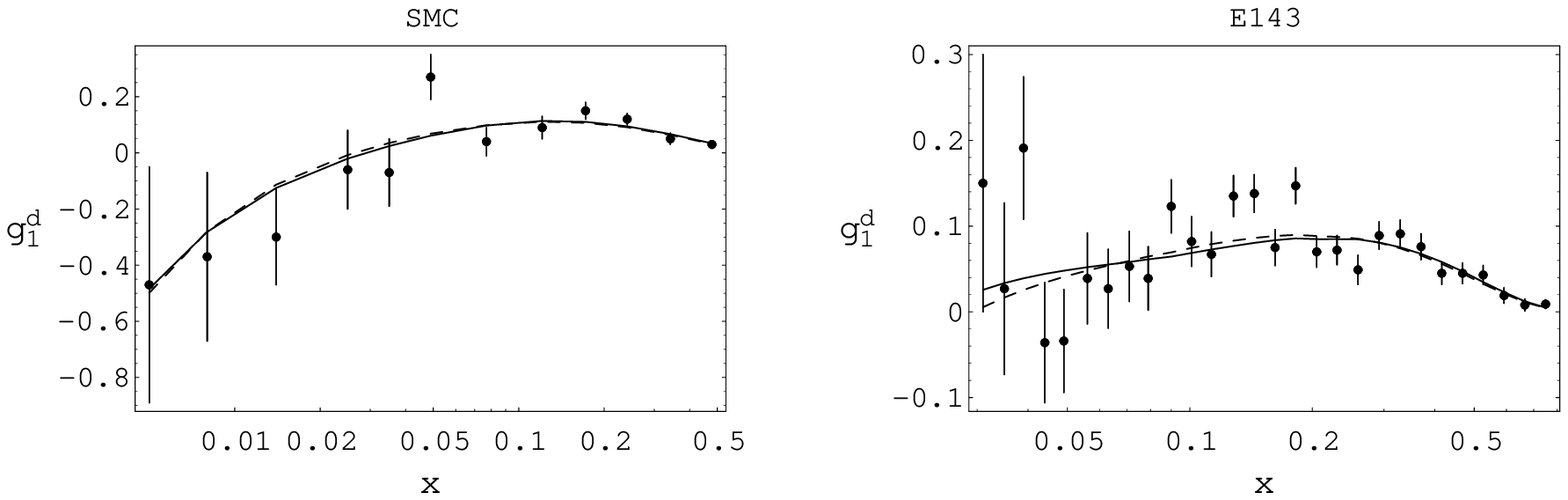}
\vspace{0.5cm}

 \noindent \hspace{-0.5cm} \epsfxsize
=400pt\epsfbox{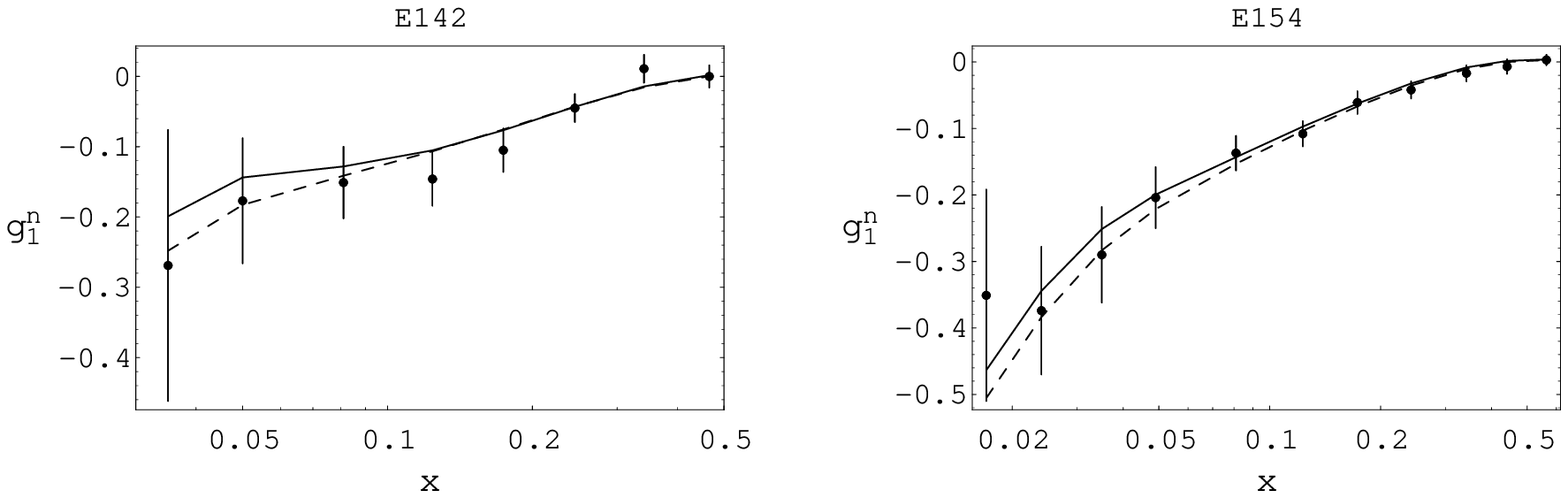} \caption{\em{The comparison of our
predictions for $g_1^N(x,Q^2)$ versus $x$ with experimental data
from different experiments. Solid curve is gotten from fit $g$,
the dashed one is calculated using the parameters of fit
$A_{1}$.}}
\end{figure}

For illustration we present in Fig.1 the distributions $\Delta u$
and $\Delta d$  for our  sets of parameters calculated from
polarised structure functions and the sample of averaged spin
asymmetries data. The corresponding values for $\Delta u_v$ and
$\Delta d_v$ differ much stronger. In our previous paper \cite{BTn} we
already presented how the values of polarised structure functions
for proton, deuteron and neutron calculated from the fits to spin
asymmetries compare with the experimental data. To see what is the
difference in the values fitted directly to the polarised
structure functions and the values calculated from the fit to spin
asymmetries we present in Fig.2 the corresponding curves in
comparison with experimental points for $g_{1}^{p}$, $g_{1}^{d}$
and $g_{1}^{n}$ at the values of $Q^2$ in corresponding
experiment. As an example we show comparison with experimental
points for  polarised structure functions $g_{1}^{p}$, $g_{1}^{d}$
from SMC from CERN and E143 from SLAC and $g_{1}^{n}$ from E142
and E154 from SLAC. There are some differences but they are not
big in comparison with experimental errors. The comparisons for
other experimental sets look very similar. It is of course not
astonishing that the fitted curves with the parameters from the
first fit are closer to experimental values.

As we already pointed out before we have not made any assumptions
about $a_8$. We got from the fits that the value of $a_8$ is near
zero very far from the experimental value and we got positive
values for $\Delta s$. The value of $a_3$ that also was not
constrained in  the fit is  close to experimental value (in the
measured region of $x$). In order to make more direct comparison
with other fits as before  we will also not fix $a_8=\Delta
u+\Delta d-2 \Delta s$ value but we will add experimental point
$a_8 = 0.58 \pm 0.1$ with enhanced (to 3$\sigma$) error. That
means we will simply add to $\chi^{2}$ corresponding to
experimental points for spin asymmetries the term connected with
experimental point from hyperon decays. The parameters of our
three new fits (called $g'$, $A'_{1}$ and $A'_{2}$ ) are now
presented in Table 2 and results in the measured region in  $0.003
\leq x \leq 1$ in the Table 2a.

\vspace{1cm}
\newpage
Table 2. {\em{The parameters of three new fits calculated  at $
Q^{2}=1\, {\rm
 GeV^{2}}$  }}
 \vspace{0.5cm}

 \hspace{-1.5cm}
\begin{tabular}{||c|c|c|c|c|c|c||} \hline
fit&$a_1$&$a_2$&$a_4$&$b_2$&$b_3$&$c_1$
\\ \hline
fit $g'$ &$0.61 $&$-6.84$&$16.83 $&$-1.86$&$0.13$&$-0.31$\\ \hline
fit $A'_{1}$ &$0.56$&$-5.51$&$14.73$&$-1.65$&$-0.21$&$-0.34$
\\  \hline
fit $A'_{2}$ & $0.50$&$-5.39$&$14.72$&$-1.98$&$0.29$&$-0.33$
\\ \hline
\end{tabular}
\vspace{0.5cm}

\hspace{3cm}
\begin{tabular}{||c|c|c|c|c|c|c||} \hline
fit&$c_2$&$c_{2s}$&$c_3$&$d_2$&$d_3$&$\chi^2/N_{DF}$
\\ \hline
fit $g'$ &$4.15 $&$-0.54 $&$-1.04 $&$-34.5$&$102.2$&$0.88$
\\ \hline fit $A'_{1}$&
$3.67$&$-0.63$&$-0.64$&$-15.8$&$43.5$&$0.80$
\\  \hline
fit $A'_{2}$ &$4.15$&$-0.56$&$-0.44$&$-14.2$&$27.6$&$0.84$
\\ \hline
\end{tabular}
\vspace{0.5cm}

There are some small changes in the parameters in comparison to the
 fits g, $A_{1}$, $A_{2}$ and the biggest
change is in $c_{2s}$ the parameter responsible for the strange
sea. Strange quark contribution is not well determined by the
polarised deep inelastic data alone and it is easy by additional
experimental point on $a_8$ from hyperon decays to shift the value
of $a_8$ from nearly zero to correct experimental value with only
small changes in non strange parton parameters. Comparing Table 1a
and Table 2a we can see what is the influence of this additional
experimental point $a_8$ on integrated parton densities for our
three fits. This additional experimental point causes shifts of
integrated parton values. As is seen from Table 2a the valence
non-strange quark distributions nearly cancel and the value of
$a_8=\Delta u +\Delta d -2\Delta s=0.58$ is built up from
relatively high sea contributions. The values of $\Gamma^{p}_{1}
$, $\Gamma^{n}_{1}$ and $a_{3}$ do not change in comparison to
previous fits. One can easily calculate from the Table 2a that
$\Delta \Sigma=0.24$ in the fits $g'$ and $A'_{1}$. \vspace{1cm}

Table 2a. {\em{The values of quark and gluon polarisations  at $
Q^{2}=1\, {\rm
 GeV^{2}}$  for three new fits (where one includes experimental
 point from hyperon decays).}}
\vspace{0.5cm}

\begin{tabular}{||c|c|c|c|c|c|c|c|c||} \hline
fit&$\Delta u $&$\Delta d $&$\Delta u_v $&$\Delta d_v $&$2\Delta
\bar{u} $&$2\Delta\bar{d}$&$2\Delta \bar{s} $&$\Delta G$
\\ \hline
fit $g'$&$0.79 $&$-0.44$&$0.47 $&$-0.60$&$0.32$&$0.16$&$-0.11
$&$0.16 $\\ \hline fit $A'_{1}$&
$0.80$&$-0.44$&$0.57$&$-0.57$&$0.23$&$0.13$&$-0.12$&$0.01$
\\ \hline
fit $A'_{2}$&
$0.80$&$-0.43$&$0.54$&$-0.62$&$0.26$&$0.19$&$-0.11$&$-0.19$
\\  \hline
\end{tabular}
\vspace{0.5cm}

 However it is not clear whether
the general conclusion that the sea contributions for quarks (
both non-strange and strange) are very big in the measured $x$
region is correct. It is specific for our model that the leading
singularity for polarised quark valence and sea contributions are
comparable. That means that splitting into valence and sea
contributions could be not well determined in our fits. That of
course could be connected with the functional form of polarised
parton densities used by us but what is important is to reproduce
the results of the experiment in the measured region. That means
that only $\Delta u$ and $\Delta d$ values can be determined using
our parametrisation from the polarised deep inelastic data and not
valence and sea contributions separately. The value of $\Delta s$
is determined only with additional $a_8$ value from hyperon
decays. In our model $\Delta G$ is small. The integrated values
$\Delta u=0.80$, $\Delta d=-0.44$ are actually not unexpected.
Similar values follow from the other models. In other fits \cite{grv,grvn} with
completely different assumptions for example when we assume the
values of $a_3$ and $a_8$ by fixing the parameters of the fitted
parton distributions (normalisation constants) and with the
assumption of $SU(3)$ symmetry for quark sea we get (using
completely different  parametrisation from that we use for
polarised parton densities):
\begin{eqnarray}
\Delta u_{v}- \Delta d_{v}&=&1.26, \nonumber \\
 \Delta
u_{v}+\Delta d_{v}&=&0.58
\end{eqnarray}

If follows that $\Delta u_v=0.92$, $\Delta d_v=-0.34$ and in order
to get $\Delta \Sigma=0.20$ \cite{grvn} we get $2\Delta
\bar{u}=2\Delta \bar{d}=2\Delta \bar{s}=-0.13$ and that means
$\Delta u=0.79$, $\delta d=-0.47$ ( the values not very different
from our values). In such models sea contribution is relatively
big and negative contrary to our model where we have at least in
the measured region big and positive sea contribution. This type
of splitting into big valence and relatively big negative sea
contribution is caused by the assumptions of the model. We have
not made such assumptions taking $a_8$ as additional experimental
point. Our solution with relatively small valence contribution and
relatively large positive sea contributions is different but the
values of $\Delta u$ and $\Delta d$ in both models are very close.
It seems that our assumptions are less restrictive. The fact that
we can get completely different splitting into valence and sea
contributions using the same experimental data shows that this
splitting is not well determined by experimental data what is more
reliable are distributions of $\Delta u$ and $\Delta d$ and their
integrated values.

We have made fits to data on polarised structure functions on
proton, neutron and deuteron targets and we have determined
polarised parton distributions. These fits are compared with our
previous fits for corresponding asymmetries (improved by usage of
recent data from E155 proton experiment in SLAC). As a check we
also have made fits to spin asymmetries for all (non averaged in
$Q^2$) data on spin asymmetries. These fits lead to very similar
results with small integrated gluon contribution. The fits were
made without inclusion of  information on $a_3$ and $a_8$ from
hyperon decays and then repeated with additional experimental
point on $a_8$. In the first case $a_8$ is close to zero and
$\Delta s$ is positive. In the second case additional experimental
point on $a_8$ changes practically  only parameters of strange
quark and causes small shifts in other parameters. The value of
$a_3$ at least in the measured region of $x$ without any
assumptions comes out very close to experimental value. It seems
that with the parametrisation used by us only $\Delta u(x)$ and
$\Delta d(x)$ are well determined (not the splitting into valence
and sea parts). Polarised strange quark distributions, gluon
distributions and also $\bar{u}-\bar{d}$ splitting are not well
determined by polarised deep inelastic experimental data.


\newpage

\begin{thebibliography}{99}
\bibitem{e80} M.J.Alguard {\em et al.}, Phys. Rev. Lett. {\bf 37}, 1261
(1976); Phys. Rev. Lett. {\bf 41}, 70 (1978);
\bibitem{e130} G.Baum {\em et al.}, Phys. Rev. Lett. {\bf 45}, 2000 (1980);
{\bf 51}, 1135 (1983);
\bibitem{e142n} E142 Collaboration, P.L.Anthony {\em et al.}, Phys. Rev.
Lett. {\bf 71}, 959 (1993); Phys. Rev. {\bf D 54}, 6620 (1996);
\bibitem{e143p} E143 Collaboration, K.Abe {\em et al.}, Phys. Rev. Lett.
{\bf 74}, 346 (1995);
\bibitem{e143d} E143 Collaboration, K.Abe {\em et al.}, Phys. Rev. Lett.
{\bf 75}, 25 (1995);
\bibitem{e154n} E154 Collaboration, K.Abe {\em et al.}, Phys. Rev. Lett.
{\bf 79}, 26 (1997); Phys. Lett. {\bf B 405}, 180 (1997);
\bibitem{e143qd} E143 Collaboration, K.Abe {\em et al.}, Phys. Lett.
{\bf B 364}, 61 (1995);
\bibitem{e143new} E143 Collaboration, K.Abe {\em et al.}, Phys. Rev.
{\bf D 58}, 112003 (1998);
\bibitem{e155} E155 Collaboration, P.L.Anthony {\em et al.}, Phys. Lett.
{\bf B 463}, 339 (1999);
\bibitem{e155p} E155 Collaboration, P.L.Anthony {\em et al.}, Phys. Lett.
{\bf B 493}, 19 (2000);
\bibitem{emc} European Muon Collaboration, J.Ashman {\em et al.},
Phys. Lett. {\bf B 206}, 364 (1988); Nucl. Phys. {\bf B 328}, 1 (1989);
\bibitem{SMCd} Spin Muon Collaboration, B.Adeva {\em et al.}, Phys. Lett.
{\bf B 302}, 533 (1993); D.Adams {\em et al.},
Phys. Lett. {\bf B 357}, 248 (1995);
\bibitem{SMCp} Spin Muon Collaboration, D.Adams {\em et al.}, Phys. Lett.
{\bf B 329}, 399 (1994); B.Adeva {\em et al.}, Phys. Lett. {\bf B 412},
414 (1997);
\bibitem{SMCqp} Spin Muon Collaboration, D.Adams {\em et al.}, Phys. Rev.
{\bf D 56}, 5330 (1997);
\bibitem{SMCqd} Spin Muon Collaboration, D.Adams {\em et al.}, Phys. Lett.
{\bf B 396}, 338 (1997);
\bibitem{SMCnew} Spin Muon Collaboration, B.Adeva {\em et al.}, Phys. Rev.
{\bf D 58}, 112001 (1998);
\bibitem{hermes} Hermes Collaboration, K.Ackerstaff {\em et al.},
Phys. Lett. {\bf B 404}, 383 (1997);
\bibitem{hernew} Hermes Collaboration, A.Airapetian {\em et al.},
Phys. Lett. {\bf B 442}, 484 (1998);
\bibitem{dg} A.V.Yefremov, O.V.Teryaev, Dubna
Report No. JIN-E2-88-287 (1988);
 G.Altarelli, G.G.Ross, Phys. Lett. {\bf B 212}, 391 (1988);
R.D.Carlitz, J.D.Collins, A.H.Mueller, Phys. Lett. {\bf B 214}, 219 (1988);
\bibitem{grv} M.Gl\"{u}ck, E.Reya, M.Stratman, W.Vogelsang,
Phys. Rev. {\bf D 53}, 4775 (1996);
\bibitem{alt0} G.Altarelli, R.D.Ball, S.Forte, G.Ridolfi, Acta Phys. Pol.
 {\bf B 29}, 1145 (1998);
 \bibitem{SMCth} Spin Muon Collaboration, B.Adeva {\em et al.}, Phys. Rev.
{\bf D 58}, 112002 (1998);
\bibitem{inni} T.Gehrmann, W.J.Stirling, Phys. Rev. {\bf D53}, 6100
(1996); G.Altarelli, R.D.Ball, S.Forte, G.Ridolfi, Nucl. Phys.
 {\bf B 496}, 337 (1997); C.Bourrely, F.Buccella, O.Pisanti, P.Santorelli,
J.Soffer, Prog. Theor. Phys. {\bf 99}, 1017 (1998); E.Leader, A.V.Sidorov,
D.B.Stamenov, Int. J. Mod. Phys. {\bf A 13}, 5573 (1998);
\bibitem{leader} E.Leader, A.V.Sidorov,
D.B.Stamenov, Phys. Rev. {\bf D58}, 114028 (1998);
\bibitem{goto} Y. Goto et al. Phys. Rev. {\bf D62}, 034017 (2000);
D. de Florian, R. Sassot  Phys. Rev. {\bf D62}, 094025 (2000);
\bibitem{BT} S.Tatur, J.Bartelski, M.Kurzela; Acta Phys. Pol.
{\bf B 31}, 647 (2000);
\bibitem{BTn} S.Tatur, J.Bartelski; Acta Phys. Pol. {\bf B 32}, 2101 (2001);
\bibitem{bt1} J.Bartelski, S.Tatur, Acta Phys. Pol. {\bf B 26},
913 (1994); J.Bartelski, S.Tatur, Zeit. f. Phys. {\bf C 71},
595 (1996); J.Bartelski, S.Tatur, Acta Phys. Pol. {\bf B 27},
911 (1996); J.Bartelski, S.Tatur Zeit. f. Phys. {\bf C 75},
477 (1997);
\bibitem{MRSTnew} A.D.Martin, W.J.Stirling, R.G.Roberts, R.S.Thorne;
Eur. Phys. J. {\bf C 4}, 463 (1998);
\bibitem{whitn}  E143 Collaboration, K.Abe {\em et al.}, Phys. Lett.
{\bf B 452}, 194 (1999);
\bibitem{ewol2} M.Gl\"{u}ck, E.Reya, A.Vogt,
Zeit. f. Phys. {\bf C 48}, 471 (1990);
\bibitem{whit}  L.W.Whitlow {\em et al.}, Phys. Lett.
{\bf B 250}, 193 (1990);
\bibitem{grvn} M.Gl\"{u}ck, E.Reya, M.Stratman, W.Vogelsang,
Phys. Rev. {\bf D 63}, 094005 (2001).
\end{thebibliography}
\end{document}